# The trigger system of the ICARUS experiment for the CNGS beam

M. Antonello[1], B. Baibussinov[2], P. Benetti[3], F. Boffelli[3], A. Bubak[11], E. Calligarich[3], S. Centro[2], A. Cesana[4], K. Cieslik[5], D. B. Cline[6], A.G. Cocco[7], A. Dabrowska[5], D. Dequal[2], A. Dermenev[8], R. Dolfini[3], A. Falcone[3], C. Farnese[2], A. Fava[2,*], A. Ferrari[9], G. Fiorillo[7], D. Gibin[2], S. Gninenko[8], A. Guglielmi[2], M. Haranczyk[5], J. Holeczek[11], M. Kirsanov[8], J. Kisiel[11], I. Kochanek[11], J. Lagoda[10], S. Mania[11], A. Menegolli[3], G. Meng[2], C. Montanari[3], M. Nicoletto[2], S. Otwinowski[6], P. Picchi[12], F. Pietropaolo[2], P. Plonski[14], A. Rappoldi[3], G.L. Raselli[3], M. Rossella[3], C. Rubbia[1,9,13], P. Sala[4], A. Scaramelli[4], E. Segreto[1], F. Sergiampietri[15], D. Stefan[4], R. Sulej[10,9], M. Szarska[5], M. Terrani[4], M. Torti[3], F. Varanini[2], S. Ventura[2], C. Vignoli[1], H. Wang[6], X. Yang[6], A. Zalewska[5], A. Zani[3], K. Zaremba[14]

(ICARUS Collaboration)

[1] *INFN - Laboratori Nazionali del Gran Sasso, Assergi, Italy*
[2] *Dipartimento di Fisica e Astronomia Università di Padova and INFN, Padova, Italy*
[3] *Dipartimento di Fisica Università di Pavia and INFN, Pavia, Italy*
[4] *Politecnico di Milano and INFN, Milano, Italy*
[5] *H. Niewodniczanski Institute of Nuclear Physics, Polish Academy of Science, Krakow, Poland*
[6] *Department of Physics and Astronomy, UCLA, Los Angeles, USA*
[7] *Dipartimento di Scienze Fisiche Università Federico II di Napoli and INFN, Napoli, Italy*
[8] *INR RAS, Moscow, Russia*
[9] *CERN, Geneva, Switzerland*
[10] *National Centre for Nuclear Research, Otwock/Swierk, Poland*
[11] *Institute of Physics, University of Silesia, Katowice, Poland*
[12] *INFN Laboratori Nazionali di Frascati, Frascati, Italy*
[13] *GSSI, L'Aquila, Italy*
[14] *Institute of Radioelectronics, Warsaw University of Technology, Warsaw, Poland*
[15] *INFN, Pisa, Italy*

*E-mail*: Angela.Fava@pd.infn.it

ABSTRACT: The ICARUS T600 detector, with its 470 tons of active mass, is the largest liquid Argon TPC ever built. Operated for three years in the LNGS underground laboratory, it has collected thousands of CNGS neutrino beam interactions and cosmic ray events with energy spanning from tens of MeV to tens of GeV, with a trigger system based on scintillation light, charge signal on TPC wires and time information (for beam related events only). The performance of trigger system in terms of efficiency, background and live-time as a function of the event energy for the CNGS data taking is presented.

KEYWORDS: Trigger concepts and systems (hardware and software), Noble liquid detectors (scintillation, ionization, double-phase), Neutrino detectors.

# Contents

## 1. Introduction

The ICARUS T600 detector installed in the underground INFN-LNGS Gran Sasso Laboratory has been the first large-mass Liquid Argon TPC (LAr-TPC) operating as a continuously sensitive general-purpose observatory [1]. The successful operation of the ICARUS T600 LAr-TPC demonstrates the enormous potential of this detection technique [2], addressing a wide physics program with the simultaneous exposure to the CNGS neutrino beam [3] and cosmic-rays.

ICARUS T600 is expected to undergo an overhauling and to be complemented with a new T150 smaller "clone" of ¼ of T600 for sterile neutrino search on a short baseline neutrino beam [4]. In addition, this LAr program may also pave the way to the ultimate realization of multi-kton liquid Argon detectors for future long baseline neutrino experiments [5],[6].

In the framework of the search for $\nu_\mu \rightarrow \nu_\tau$ appearance and $\nu_\mu \rightarrow \nu_e$ oscillation driven by an LSND-like anomaly, ICARUS T600 has detected thousands of CNGS neutrino interactions clustered in a $10 \div 35$ GeV energy window. The additional recording of cosmic-ray induced events has provided a large data sample for addressing several other physics topics. In particular, the search for nucleon decay focuses on localized events with energy deposition < 1 GeV, while atmospheric neutrino interactions range from few hundreds of MeV to tens of GeV.

The wide set of event types, spanning few orders of magnitude in energy deposition and with topologies significantly different from each other, is challenging also for the trigger system. The ICARUS T600 detector relies on its self-triggering capability, using for the first time both scintillation light and ionization signals produced by charged particles in LAr. The trigger for the CNGS neutrino events has been set up including also the prediction of the CERN SPS proton beam extraction time provided by an "early warning" sent from CERN to LNGS. The successful operation of the ICARUS

T600 trigger system allowed the collection of CNGS events with high reliability, efficiency and live-time, representing a robust baseline for future developments in multi-kton LAr-TPC neutrino detectors.

This paper is devoted to the description of the solutions adopted for the trigger architecture of the ICARUS T600 experiment. The performance of the trigger system for the data taking with the CNGS neutrino beam in terms of rates, efficiency, live-time and background will also be presented.

## 2. Scintillation and ionization signals in ICARUS T600

The ICARUS T600 detector consists of a large cryostat split into two identical, adjacent modules filled with about 760 tons of ultra-pure liquid Argon. A detailed description can be found elsewhere [1],[7]. Each module houses two TPCs with 1.5 m maximum drift path, sharing a common cathode made of punched inox sheets with 58 % transparency to light. A uniform electric field ($E_{drift}$ = 500 V/cm) drifts ionization electrons with $v_D$ ~ 1.6 mm/µs velocity towards the anode, consisting in three wire arrays that guarantee a stereoscopic event reconstruction (Figure 1). A total of 53248 wires are deployed, with a 3 mm pitch, oriented on each plane at a different angle (0°, +60°, -60°) with respect to the horizontal direction. By appropriate voltage biasing, the first two wire planes (Induction1 and Induction2 planes) provide signals in a non-destructive way; finally the ionization charge is collected and measured on the last plane (Collection).

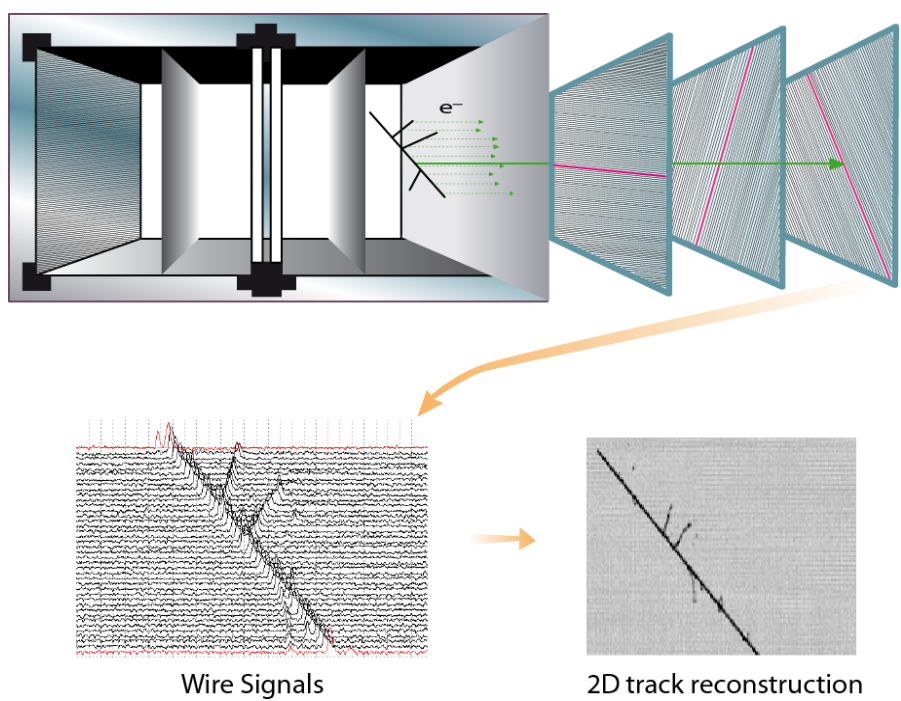

**Figure 1. Working principle of ionization charge detection in ICARUS T600.**

Charged particles deposit energy in liquid Argon mainly by excitation and ionization of Ar atoms, leading to scintillation light emission and free electron production, respectively. Additional scintillation light comes from the recombination of electron-ion pairs, which is inversely proportional to the strength of the electric field applied to the detector active volume. As a consequence, free-electron yield rises with the field value while photon yield decreases. In both cases saturation occurs for $E_{drift}$ > 10 kV/cm. At the nominal drift field applied in ICARUS T600, approximately the same amount of photons (~ 4000 γ/mm) and free electrons (~ 5000 ion-electron couples per mm) are produced for minimum ionizing particles [8].

Scintillation light emission in LAr is due to the radiative decay of excited molecules ($Ar_2^*$) produced by ionizing particles, releasing monochromatic VUV photons (λ ~ 128 nm) in transitions from the lowest excited molecular state to the dissociative ground state [9]. A fast (τ ~ 6 ns decay time) and a slow (τ ~ 1.6 μs) components are emitted; their relative intensity depends on dE/dx, ranging from 1:3 in case of minimum ionizing particles up to 3:1 in case of alpha particles. This isotropic light signal propagates with negligible attenuation throughout each TPC volume. Indeed, LAr is fully transparent to his own scintillation light, with measured attenuation length in excess of several tens of meters and Rayleigh-scattering length of the order of 1 m. Because of their short wavelength the scintillation photons are absorbed by all detector materials without reflection. In ICARUS T600 direct light is detected by Photo-Multiplier Tubes (PMTs) immersed in the LAr, for absolute timing of events and triggering purposes.

Free electrons from the ionization process are exploited to retrieve a faithful 3D image of each event (Figure 1) with a remarkable resolution of ~ 1 $mm^3$, by combining the coordinates on the three wire planes with the drift time information. The excellent level of LAr purity obtained in the whole detector and maintained during the entire data taking period in steady conditions, with free electron lifetime exceeding 5 ms [1], ensured a very low attenuation of the free electron yield (16 %, at the maximum drift distance). In ICARUS T600 the charge deposition signal on TPC wires was also used for self-triggering purposes on localized events.

## 3. The ICARUS T600 PMT system

The LAr scintillation light trigger system is based on the detection and exploitation of both scintillation light components by means of PMTs directly immersed in the liquid. VUV photons are converted to visible light by means of an appropriate wavelength shifter. Each PMT response was integrated with a time constant >> 1.6 μs decay time of the slow component of light. Four trigger signals, obtained as the sum of the PMT signals from each TPC, were discriminated and sent to the detector Trigger Manager (Section 6).

### 3.1 The PMT system set-up

The T600 detector PMT system set-up was realized according to the results from dedicated R&D activities on the LAr scintillation light detection [10]. The adopted solution is based on the large surface photomultiplier 9357FLA Electron Tube, a 12-stage dynode PMT with hemispherical glass window 200 mm (8") diameter, manufactured to work at cryogenic temperatures [11]. The main physical and electric characteristics of the device can be summarized as: 300 ÷ 500 nm spectral response; 5 ns rise time and 8 ns FWHM; $5 \times 10^7$ maximum gain with 18 % maximum quantum efficiency (blue) with Platinum (Pt) under-layer.

The PMT sensitivity to VUV photons (128 nm) was achieved by coating the glass window with Tetra-Phenyl-Butadiene (TPB), which acts as fluorescent wavelength shifter to the PMT sensitive spectrum. The 0.2 mg/$cm^2$ TPB coating thickness on sand-blasted glass guaranteed a conversion efficiency better than 90% and good adhesion after immersion in LAr, resulting in a PMT response with ~ 4% overall quantum efficiency [12].

The PMTs were located in the 30 cm space behind the wire planes of each TPC, at 5 mm distance from the Collection wires, with a dedicated sustaining structure especially designed to compensate the thermal stresses occurring during the cooling of the T600 cryostat (Figure 2-left).

Three rows of 9 PMTs, spaced by 2 m, found place in the East module behind each wire chamber for a total amount of 27 + 27 photo-devices. With respect to the TPC vertical coordinate, whose origin is set at the ground floor of the LNGS Hall B, the three rows were placed in central (391 cm), top (489 cm) and bottom (293 cm) positions, with the central one shifted by 86 cm along the longitudinal direction. This layout maximizes the coverage of PMT system getting a uniform response to particles interacting in the detector (Figure 2-right).

In the West module only the two central rows were deployed; two additional PMTs were placed in the top and bottom positions in the Right chamber at the center of the longitudinal direction, for an overall amount of 20 PMTs (Figure 2-right).

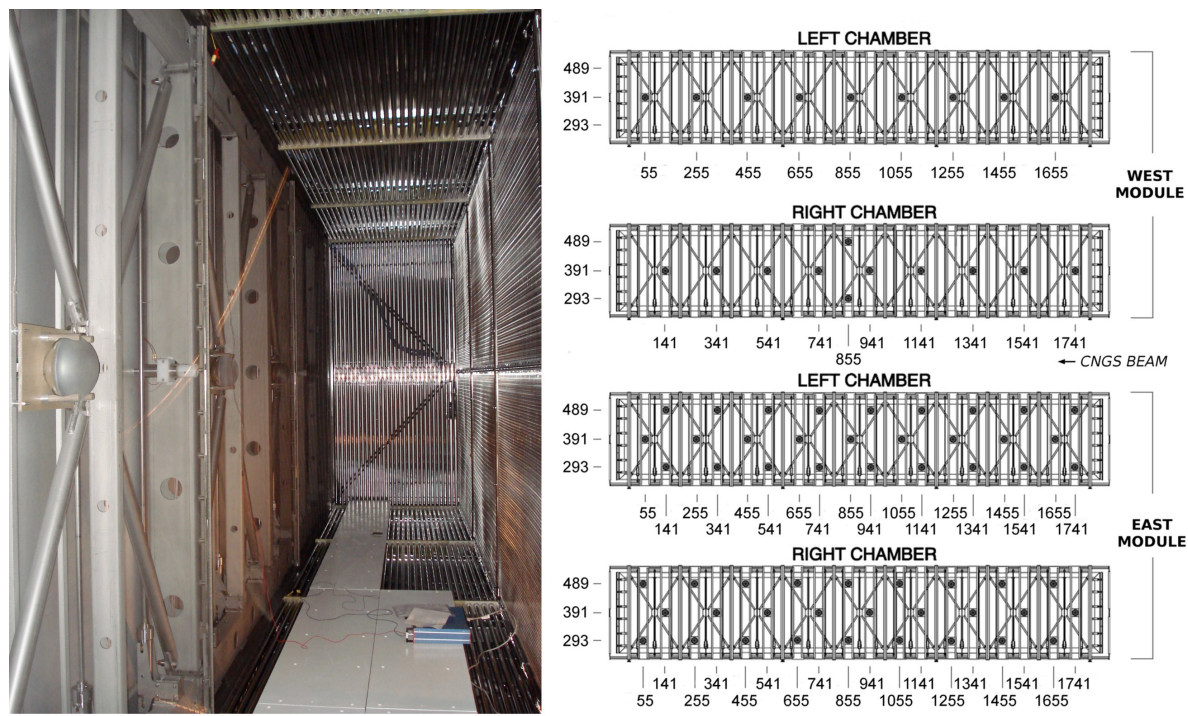

**Figure 2. Left: inside picture of one TPC chamber, with a few PMTs clearly visible together with their sustaining structure. Right: PMT's deployment in the two ICARUS T600 cryostats. PMT coordinates in cm are referred to a reference frame in Hall B whose origin is set at the ground floor (vertical axis, x), at the center of the two modules (drift coordinate, y) and at the downstream end of the wire chambers for the longitudinal direction (z) along the CNGS beam line.**

### 3.2 PMT electronics

The electronic scheme associated to each PMT is shown in Figure 3. A voltage divider, designed to work at low temperatures and welded directly on the PMT output leads, was internally connected through 8 m long RG316 cables to dedicated HV feedthroughs on the top of the detector. A single 25 m long RG58 cable was used to provide each PMT with the proper power supply and to pick-up the anode signal. For this purpose, custom circuits were realized hosting the HV distribution, the decoupling boards and a custom-made low-noise integrating preamplifier.

The integration time constant, initially adjusted to collect the fast scintillation light component only, was increased in 2012 to integrate the PMT signal over ~ 30 μs, to profit also of the slow component, with a gain of 5 mV/pC. This optimization allowed to improve the performance of the trigger efficiency for low energy events. However, due to an incorrect implementation of the voltage divider in the PMT circuitry of the West module, a 2/3 overall signal loss was observed in comparison to the East module. Since July 2012 a new HV box, with slightly different electronic components to amplify and shape the PMT signals, was installed in the Left chamber of the East module, resulting in higher gain values than in the other chambers.

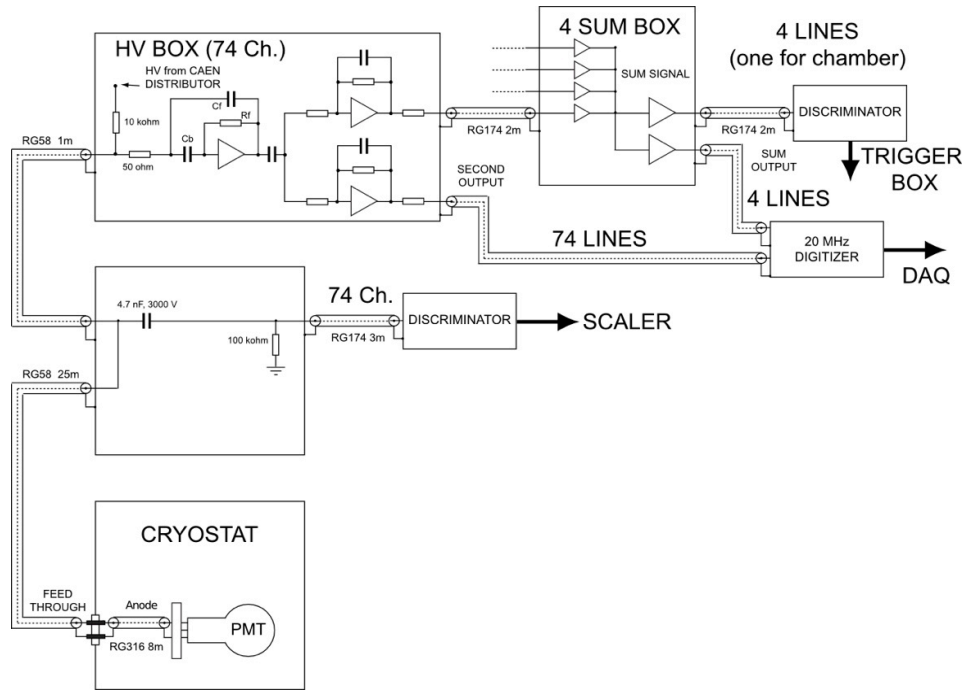

**Figure 3. PMT electronics layout.**

The PMT gains were equalized at the LAr temperature to about $10^6$, corresponding to a Single Electron Response (SER) of ~ 0.2 pC, obtained by measuring each PMT response to single photon excitation through a calibrated charge preamplifier. The electronic chain gain and linearity were determined by means of charge calibrated test pulses. The resulting overall gain was ~ 1 mV/phe (photoelectron) at the preamplifiers output, with a ~ 5% stability over the entire data taking. For each device the corresponding SER counting rate was ~ $10^4$ Hz, due to PMT noise and to single photons from low energy radioactivity in LAr, mainly $^{39}$Ar.

Preamplifier outputs were connected to 4 analogue adder electronic circuits, which provide the linear sum of the input signals with adjustable gain, to account for the different number of PMT deployed in the two modules. The overall gains, measured using test pulses, resulted to be 0.38 mV/phe and 0.12 mV/phe for the West and East module chambers, respectively.

The discrimination threshold values and the corresponding counting rates associated to the four PMT linear adders were:
- West module Left (1L)      25±1 mV    65±3 phe                                    130 Hz
- West module Right (1R)     25±1 mV    65±3 phe                                    75 Hz
- East module Left (2L)      25±1 mV    200±8 (120±5) phe (since July 2012)    100 (90) Hz
- East module Right (2R)     35±1 mV    300±12 phe                                  150 Hz.

Since July 2012, for each triggered event the PMT waveforms (Figure 4) were sampled and recorded by 19 digitizers CAEN-V791PM (10-bit ADC, 1 mV/ADC count and 20 MHz sampling rate), providing a useful tool for measuring the PMT trigger efficiency. The adopted DAQ settings allowed recording PMT waveforms without saturation for signal amplitudes below ~ 750 mV. The PMT trigger electronics was completed with a monitoring system of the single PMT counting rate.

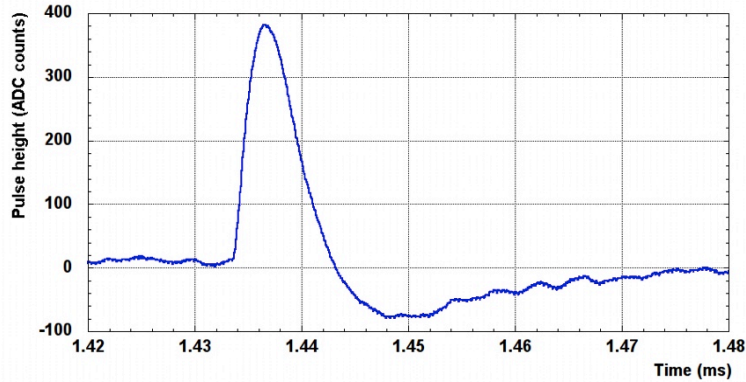

**Figure 4. Example of recorded PMT waveform after integration of fast and slow light components.**

### 3.3 PMT system performance

The angular acceptance of the single PMT in terms of collected light has been evaluated with a sample of 440 almost vertical ($\theta < 10°$) cosmic muons, spanning all the detector length with a deposited energy peaked around 750 MeV with a small width distribution.

The number of fired devices for each muon track has been determined through the distribution of the signal pulse-height as a function of the longitudinal distance of the track from the PMT in the Right Chamber of West module (Figure 5). Photons emitted from tracks more distant than the 2 m PMT spacing are detected by single PMTs, guaranteeing an almost full coverage of the LAr active volume. Signals from tracks crossing the adjacent chamber are also recorded, although with a reduced intensity due to the 58 % cathode transparency. Because of the large amount of light production, the pulse-height of the sum signals from the PMT walls exceeds the dynamic ADC range (~ 2000/6000 phe for the West/East module) for cosmic tracks crossing the TPC chamber close to the anode (Figure 6).

The efficiency of each PMT sum signal in detecting charged tracks $\varepsilon_{PMT-sum}$ has been evaluated on a sample of cosmic muons crossing a single TPC, recorded in 2012 with a "minimum bias" trigger based on the TPC wire signal. Events were triggered with the S-Daedalus (see Section 4), requiring signal from at least 12 over 16 consecutive Collection wires, corresponding to ~ 4 cm track length. The efficiency $\varepsilon_{PMT-sum}$, determined looking for events with PMT sum signal above threshold, has been studied as a function of deposited energy $E_{Dep}$, track distance from PMT wall and position of the track along the detector longitudinal axis. As expected, it increases with the energy deposition (Figure 7-top), reaching 90% and 100% values in the East module for $E_{Dep} > 300$ MeV, while slightly lower values are reached in the West module due to the smaller number of deployed PMTs. The rise of the detection efficiency with the track distance from the PMT wall in the West module (Figure 7-center) is due to the poor PMT solid angle coverage for tracks close to wires. The detection efficiency is almost optimal along all the 20 m detector length, except for the downstream region of the Right chamber in the West module where 2 PMTs were switched off (Figure 7-bottom and Figure 8). The associated systematic error ranges from 3.5(2)% in the lowest energy bin for the West(East) module down to less than 1% in the highest energy bin for both modules, according to the overall 6.4% uncertainty in the threshold level obtained combining the stability of each PMT gain with the tolerance of the PMT sum signal discriminator.

A remarkable stability of the PMT trigger system detection efficiency, well within the measurement uncertainty, has been verified by comparing different data subsets of the 2012 run.

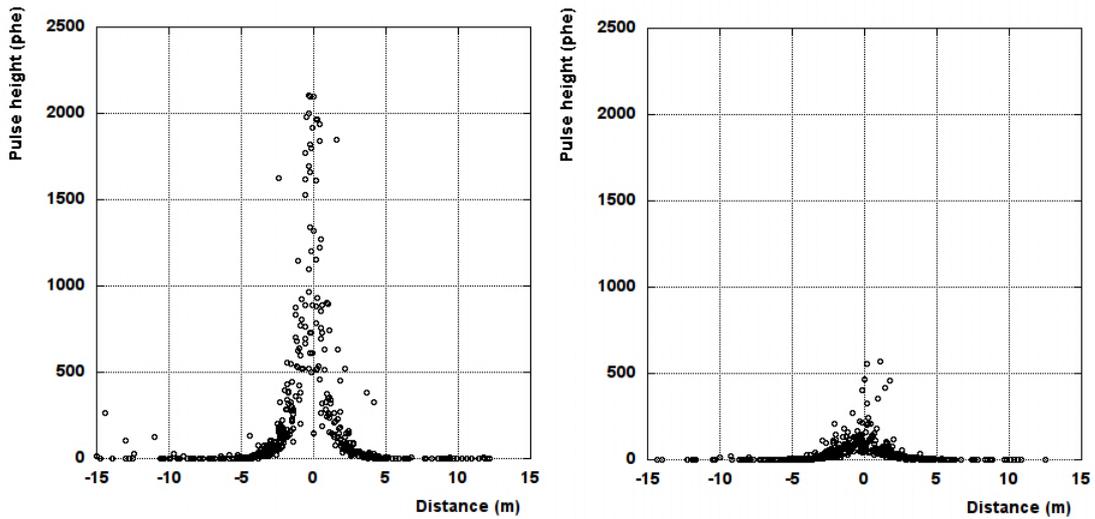

**Figure 5.** Single PMT pulse-heights in the Right Chamber of the West module as a function of the longitudinal distance of vertical tracks from the PMT when the cosmic muon is crossing the Right (left) and Left (right) TPC.

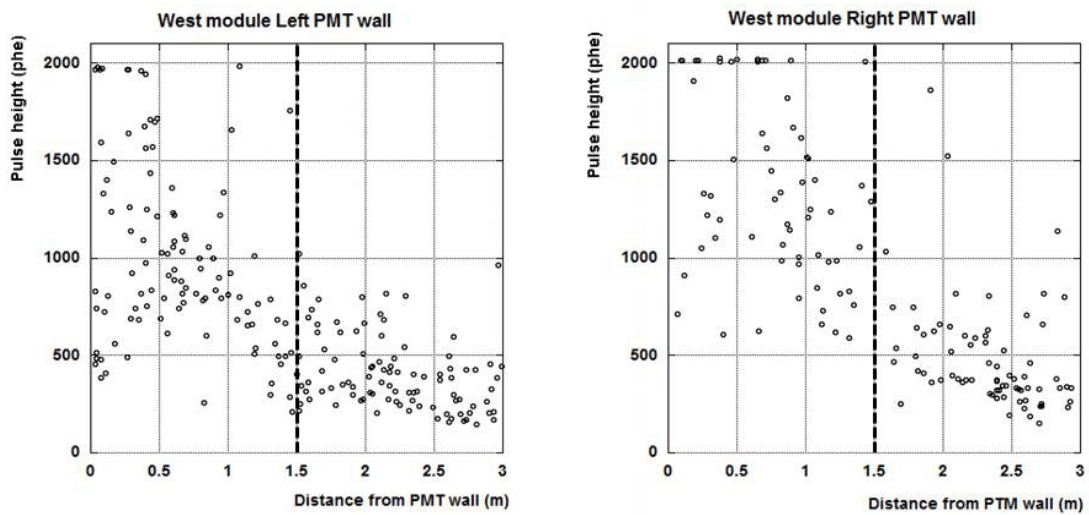

**Figure 6.** PMT sum signals in the West module Left (left) and Right (right) TPCs as a function of the track distance from the PMT wall. Dashed lines indicate the cathode position. The pulse-heights of tracks crossing the adjacent chamber have been increased according to the 58 % transparency of the cathode. Few events saturate the PMT scale at 2000 phe value.

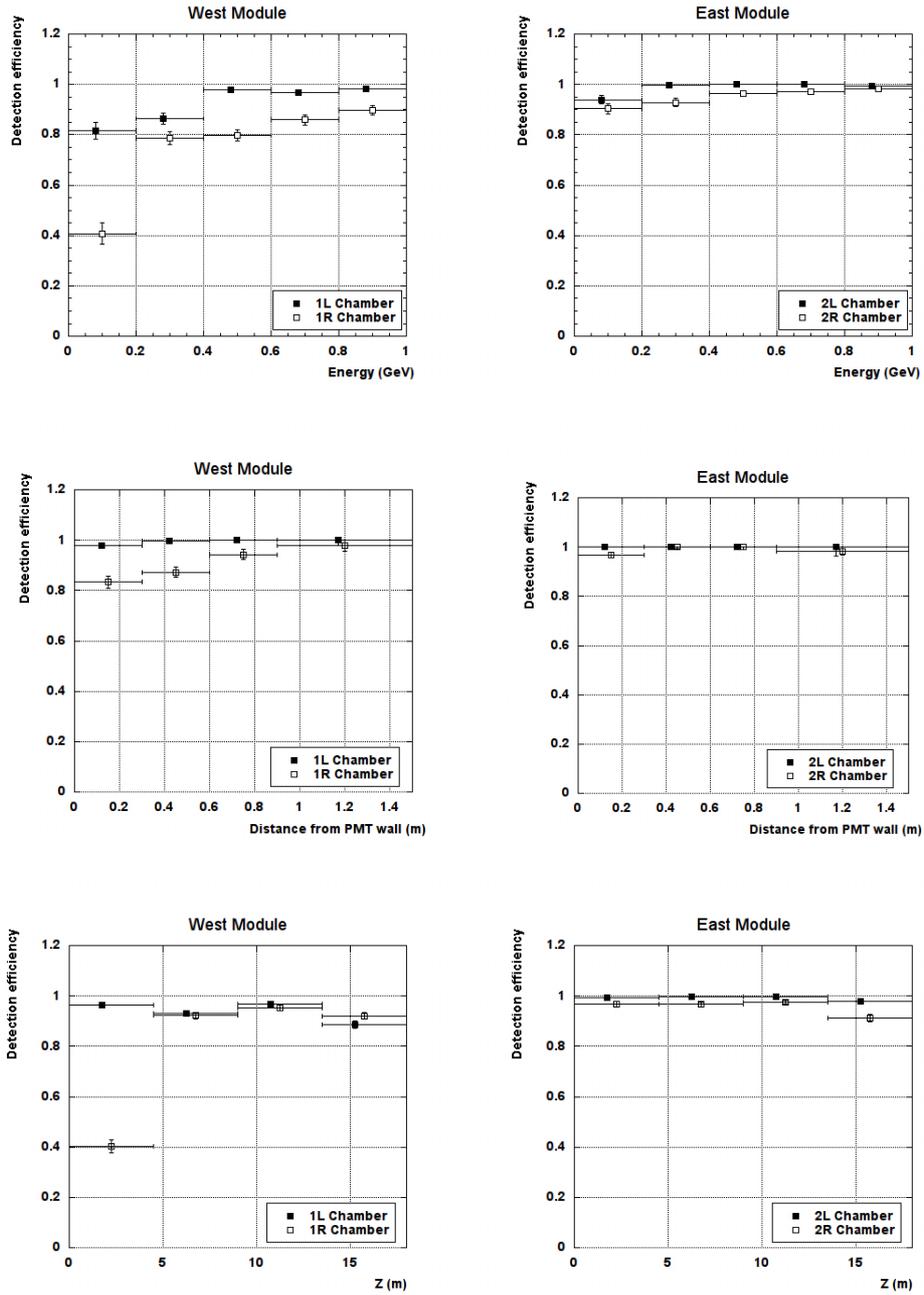

**Figure 7.** PMT sum signal efficiencies of each TPC chamber in the West (left) and East (right) modules, as a function of the energy deposited by cosmic muons (top), the track distance from the PMT wall (center) and the track longitudinal position (bottom). The represented error bars account only for the statistical error.

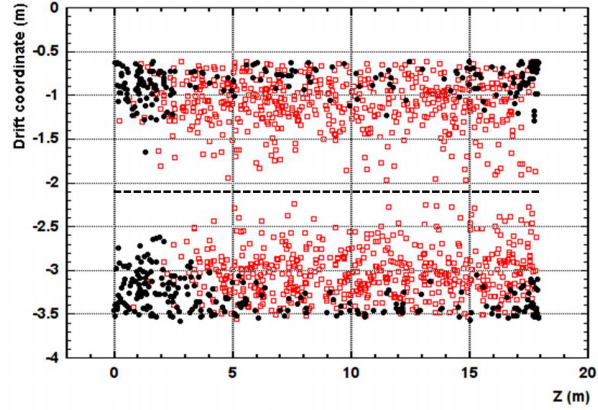

**Figure 8.** Spatial distribution of cosmic muon tracks in the West module as triggered by PMTs (red squares) and below the PMT detection threshold (black circles, see Chapter 4). PMT system inefficiency is mostly localized in the regions where PMTs were switched off.

The comparison of the pulse-height of the PMT sum signals with the corresponding energy deposited by cosmic muons allowed estimating the quantum efficiency of the light collection system (PMT + wavelength shifter). For each PMT wall only tracks crossing the chamber beyond the cathode have been considered, in order to avoid dealing with saturated ADC. The resulting scintillation light yield, normalized to the fraction of solid angle ($\Delta\Omega/4\pi$) covered by the PMTs and accounting for systematic uncertainties coming from the calibration chain procedure, were:

- West module Left (1L)   $[0.37 \pm 0.01 \text{ (stat)} \pm 0.04 \text{ (sys)}] \times 10^6$ phe/GeV
- West module Right (1R)  $[0.45 \pm 0.01 \text{ (stat)} \pm 0.05 \text{ (sys)}] \times 10^6$ phe/GeV
- East module Left (2L)   $[1.18 \pm 0.03 \text{ (stat)} \pm 0.12 \text{ (sys)}] \times 10^6$ phe/GeV
- East module Right (2R)  $[1.18 \pm 0.02 \text{ (stat)} \pm 0.12 \text{ (sys)}] \times 10^6$ phe/GeV

As expected (see Section 3.2), the collected light signal in the West module is about one third of the East one. Accounting for the $(24.0 \pm 2.4) \times 10^6\, \gamma/\text{GeV}$ photon production with $E_{drift}$ = 500 V/cm [13],[14] a PMT quantum efficiency QE = $[4.9 \pm 0.1 \text{ (stat)} \pm 0.7 \text{ (sys)}]$ % has been determined for the East module, roughly in agreement with the laboratory measurements performed before the PMT deployment in the T600 [11].

## 4. The ICARUS T600 S-Daedalus system

A second source of triggers has been gradually implemented, starting from the end of 2011, to increase the efficiency of the ICARUS T600 detector for low energy events down to few MeV and to provide a reference for measuring the PMT trigger performance. It is based on a new algorithm that detects the ionizing tracks through a digital filter of each TPC wire signal; this allows triggering on charge deposition, while the PMT waveforms are used to extract the incident time = 0 information.

### 4.1 Charge signal detection on each TPC wire

A minimum ionizing particle in LAr produces ~ 5000 electrons per mm, which corresponds to a collected charge per wire of ~ 15000 electrons due to the 3 mm wire pitch. After the amplification and shaping by front-end electronics, the wire waveform exhibits a ~ 20÷30 time-samples wide peak (1 time-sample: 400 ns) with ~ 15 ADC counts height (1 ADC count ~ 1000 electrons). The signal is also affected by up to 15 ADC counts low frequency baseline fluctuation and ~ 5 time-samples wide spikes with ~ 3 ADC counts height with respect to the local baseline. This prevents an efficient identification of the single hit signal with simple threshold discrimination (Figure 9-top).

A dedicated algorithm based on Double Rebinning Sliding Windows (DR-slw) has been developed to filter out these noise components while preserving the original signal amplitude (Figure 9-bottom). A detailed description can be found elsewhere [15]. As an additional requirement, a minimum 2 time-samples duration of the signal above threshold ($Q_{thr}$) has been introduced to generate a gate (PEAK signal), used to perform majority logics among adjacent wires. The gate width is adjustable in a $25 \div 125$ μs range to prevent loss of efficiency correlated to the non-synchronous arrival time of the signals in case of tracks inclined with respect to the wire plane. A specific veto has been added to inhibit the PEAK signals generated by PMT electrostatic induction, mimicking ionization signals on Collection wires.

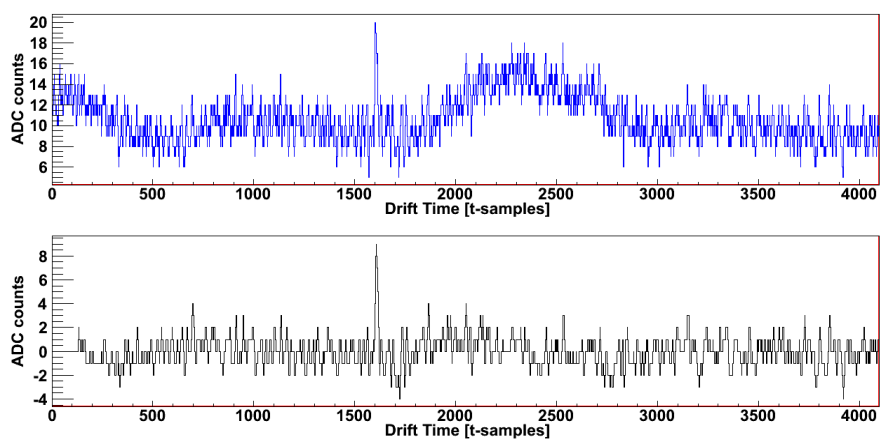

**Figure 9. Top: typical waveform of a Collection wire signal, with both high and low frequency noise components. Bottom: waveform of the same signal after the application of the DR-slw algorithm.**

According to previous laboratory tests [15], the DR-slw filter algorithm has been implemented on a Xilinx Spartan 6 XC6SLX16 FPGA installed in a dedicated piggyback, called S-Daedalus (SD), to fit in all the existing CAEN-V789 digital boards connected to TPC wires (Figure 10). Each FPGA serves two groups of 16 wires.

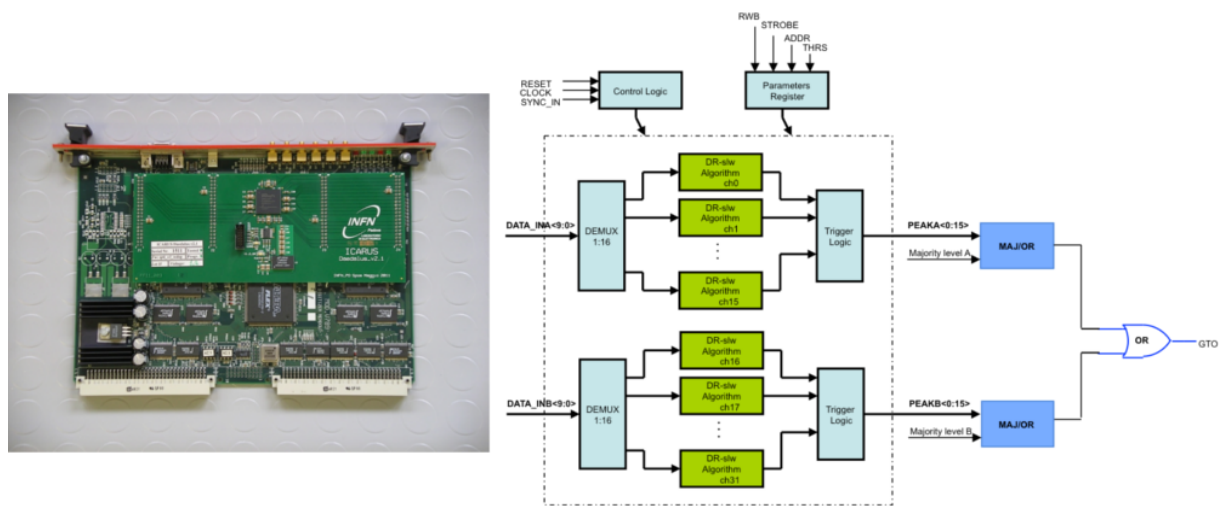

**Figure 10. Picture of the ICARUS digital read-out board (CAEN-V789) hosting the S-Daedalus piggyback board (left) and block diagram of the algorithm implemented in each FPGA (right).**

To extract a trigger signal, each set of PEAKs from 16 adjacent wires is processed by a majority stage, in order to further reduce the rate of fake triggers while preserving full efficiency even in the identification of small localized events. The logical OR of the two majorities in the same digital board generates a "Global Trigger Output" (GTO) signal, which can be used to build more complex trigger patterns.

The 18 GTO signals provided by the digital boards in each crate (576 channels) are further processed in another module housed in the same crate, in order to reduce the complexity of the global trigger algorithm and to limit the I/O and cables. This module performs the logic OR and the Majority on two independent sets of 9 GTO each, with the possibility of setting the majority level trough a dipswitch placed on the front panel. Except for few crates housing all boards of the same view, 2 GTO OR/MAJ signals come out of each crate, respectively for Collection view and for Induction2. The block diagram of this logic is shown in Figure 11. From each chamber 48 GTO OR/MAJ signals are delivered to the Trigger Manager, which exploits them to trigger on the charge deposition.

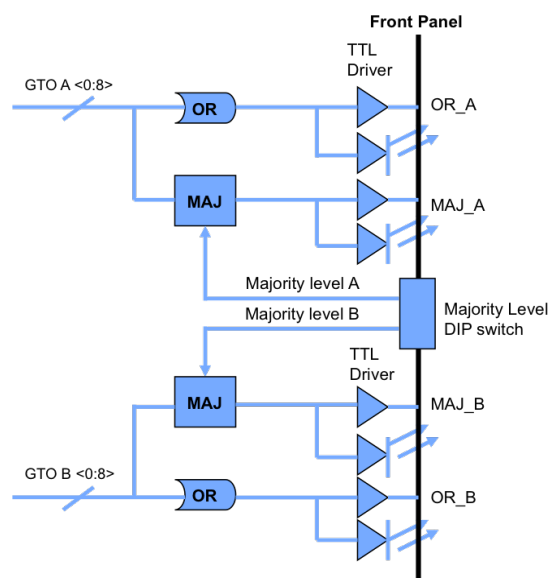

**Figure 11. Block diagram of the GTO OR/MAJ board.**

**4.2 S-Daedalus system performance**

The S-Daedalus board (SD) had been successfully tested with cosmic rays during 2009 at Laboratori Nazionali di Legnaro (LNL) with the 30 liters LAr-TPC Icarino test facility [15]. An efficiency exceeding 95 % on single hit detection and fake rejection power smaller than $10^{-3}$ for a threshold $Q_{thr} = 7$ ADC counts were achieved.

The performance of the SD system installed in ICARUS T600 was determined analyzing a sample of vertical cosmic muons collected during 2011. The single GTO detection capability, i.e. scaling up from a wire-per-wire to a board-per-board approach, has been studied for all installed SDs excluding few noisy chips (~ 3 %) whose counting rate was in excess of 30 mHz (Figure 12).

For the boards whose wires are completely crossed by the ionizing track, a detection efficiency $\varepsilon_{SD} \sim 99$ % of the single GTO in Collection view was obtained for the hit threshold $Q_{thr}$ up to 8 ADC counts requiring a majority level MAJ up to 8 over 16 wires (Figure 13). The fake signal rate is less than 1 Hz for MAJ = 8 and $Q_{thr} = 8$ ADC counts. The corresponding performance is slightly worse in Induction2 view where $\varepsilon_{SD} > 95$ % for $Q_{thr} = 6$ ADC counts with a majority level MAJ up to 6 over 16 wires; a fake rate below 1 Hz is obtained only for MAJ > 9.

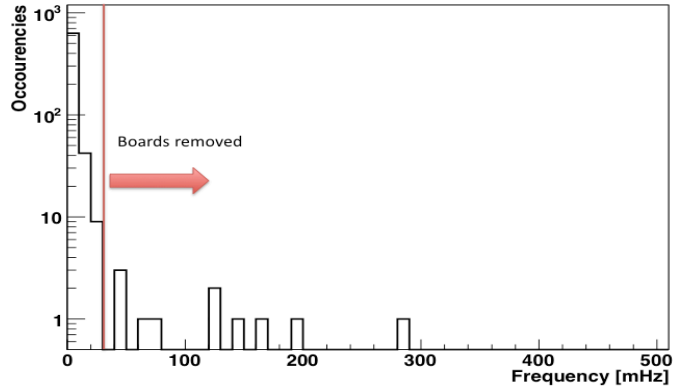

**Figure 12.** Distribution of the single GTO detection rate for all the ICARUS T600 boards. The bulk of the distribution is below 30 mHz; the few SD chips exceeding this limit have been excluded from the analysis.

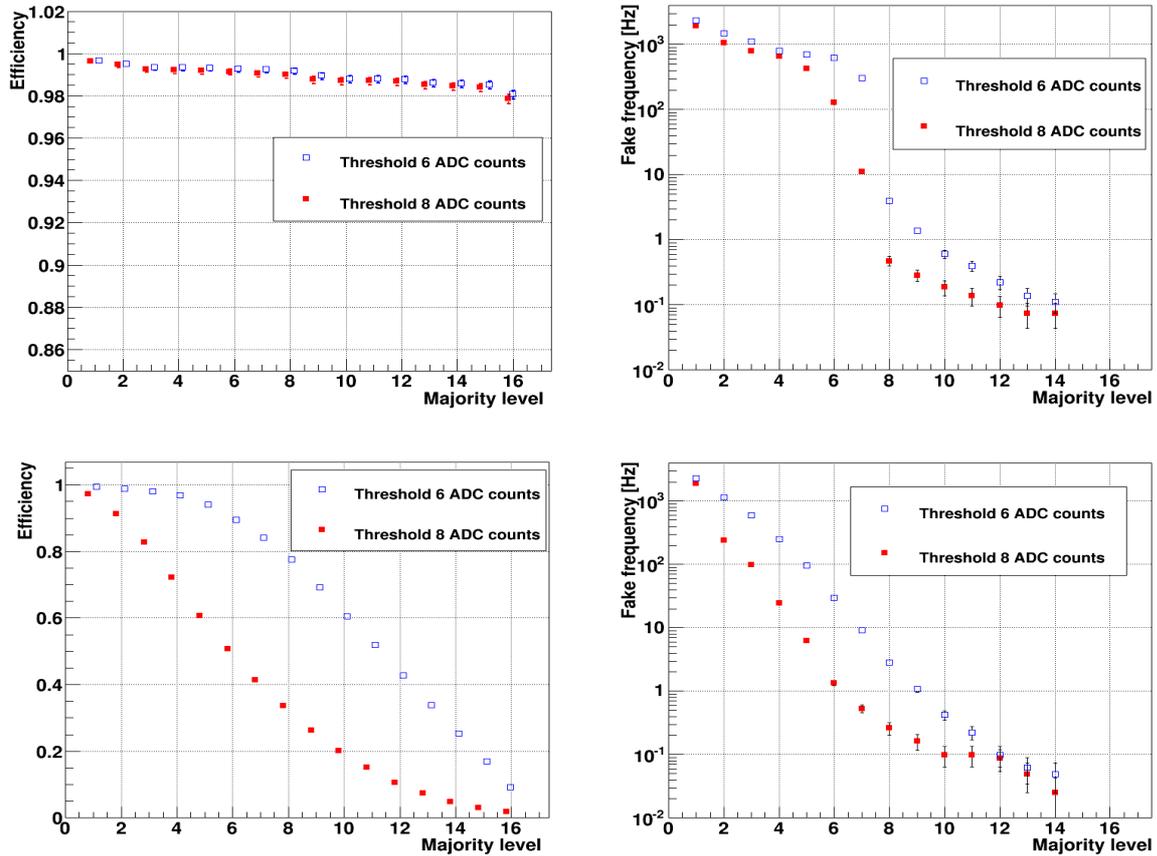

**Figure 13.** Top. Single GTO detection efficiency for vertical muon tracks (left) and corresponding rate of fake triggers (right) in Collection view as a function of the majority level for two different threshold values. The drop of the fake rate around MAJ = 7 is due to the exclusion of the noisy boards. Bottom. The corresponding GTO detection efficiency and fake rate as measured in Induction2 view. No fake triggers have been recorded for majority MAJ > 14 in both Collection and Induction2 views.

## 5. T600 synchronization with the CERN-SPS proton extraction

The synchronization with the CNGS neutrino beam was achieved by means of a common GPS time base, shared at the CERN and LNGS sites, and by a message exchange through an IP connection, which allows generating on-line a CNGS-gate signal in coincidence with the arrival of neutrinos.

At the CERN site, before each SPS proton extraction, an "Early Warning" (EW) message, containing the time predicted for the forthcoming proton extraction with order of nanosecond resolution, was sent via an UDP packet. The accuracy of this prediction has been monitored for few months during the 2011 data taking and on the whole 2012 run by comparing the EW time prediction with the actual time of extraction at a Beam Current Transformer (BCT) located in the SPS proton beam line. The jitter between the predicted and the actual proton extraction time was limited to 20 μs with a small fraction of events (< 0.1 %) exhibiting an additional 150 μs delay.

The EW packet is received at LNGS ~ 60 ms before the arrival of the neutrinos. The arrival time of the neutrino beam is calculated, and compared with the time given by a local 40 MHz oscillator disciplined by a GPS signal, sent from the LNGS external laboratory via an 8 km optical fiber. The precision of the 40 MHz clock locking to the reference GPS like signal has been measured to be better than 50 ns, while the difference of the two GPS time-base (at CERN and at LNGS) results in a < 200 ns additional jitter.

A 60 μs width CNGS-gate signal has been opened in correspondence to each predicted neutrino spill arrival, completely covering the 10.5 μs proton time extraction, to cope with all the uncertainties, dominated by the 20 μs on the prediction of the proton extraction time. The < 0.1% spills with a wrong predicted time have been discharged.

An additional 1.8% (3.8%) inefficiency in time synchronization during 2011 (2012) came from missed or late reception of the EW packets at the LNGS site (due to failures in CERN to LNGS communication), preventing the CNGS-gate opening.

## 6. Trigger Manager: description and performance

The Trigger Manager, built in a commercial National Instrument PXI crate, handled the different trigger sources (Figure 14): scintillation light collected by PMTs, timing synchronization with the CNGS extractions, charge signal collected on wires (GTO OR/MAJ) and test pulses for calibration. Furthermore, it has been programmed with a multi-veto configuration in order to assign sequential orders of priority to the different trigger sources.

The system consists of a Real Time (RT) controller (PXIe-8130) and two FPGA boards (PXI-7813R and PXI-7833). The RT controller implements all the features that imply communication with external devices, such as the DAQ process or the EW reception. Communication with the DAQ is implemented in handshake between the DAQ main process and the trigger manager. The RT controller also monitors the number of available buffers in the digital boards and prevents the generation of new triggers in case all the buffers are full. The maximum number of buffers available for full drift recording is 8. The DAQ throughput, for full drift event recoding, is limited to 0.8 Hz mainly because of the adopted VME architecture. The FPGA boards implement time critical processes, like the synchronization with the LNGS time, the opening of CNGS gate and the time stamp of each trigger. They also keep record of the trigger source and the trigger mask, monitor trigger rates from each source and control the overall system stability.

A dedicated database is devoted to log, for each event, trigger source and mask, timestamp, CNGS tag, trigger frequency and other information regarding the status of the acquisition, such as the number of occupied buffer, dead time and DAQ building time.

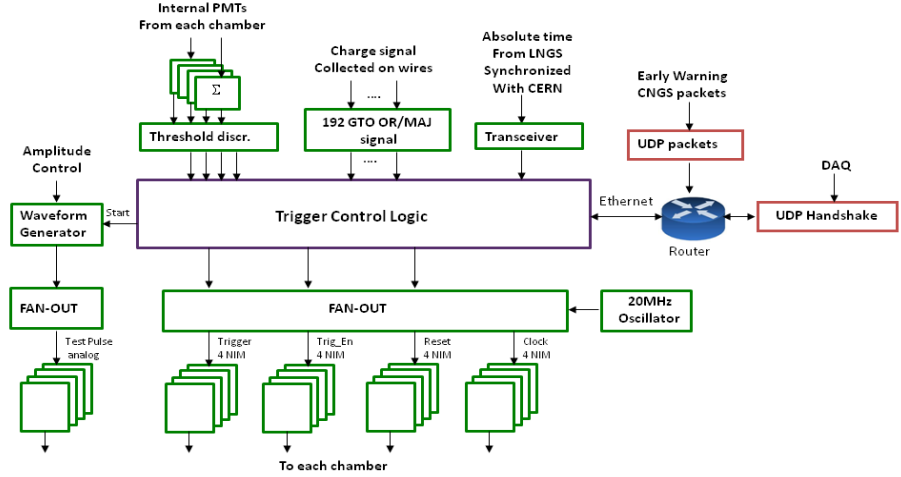

**Figure 14.** Block diagram of the ICARUS T600 Trigger Manager.

## 7. The CNGS neutrino event trigger

The CNGS neutrino beam is an almost pure $\nu_\mu$ beam with an expected average energy of 17.4 GeV and ~ 0.8 % $\nu_e$ contamination extending up to 60 GeV [3]. The anti-$\nu_\mu$ and anti-$\nu_e$ components are expected below 2% and 0.2% respectively. A $10^4$ $\nu_e$ CC, $\nu$ NC and $\nu_\mu$ CC MonteCarlo sample has been generated with the FLUKA package [16] reproducing the spectrum and direction of the CNGS neutrino beam at the Gran Sasso site and accounting for particle production and propagation in LAr. A detailed description can be found elsewhere [17]. Events due to CNGS neutrino beam interactions are expected to be characterized by significantly different topologies and energy depositions in the detector (**Figure 15**):

1) $\nu_\mu$ CC events have ~ 10 GeV visible energy and are usually characterized by a long muon track exiting from the neutrino interaction vertex and crossing most part of the detector;

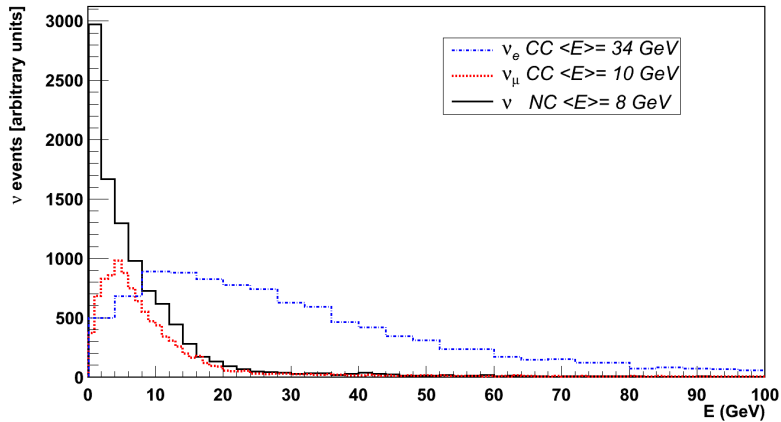

**Figure 15.** Expected deposited energy spectrum in the 0-100 GeV range, for a $10^4$ $\nu_e$ CC (blue), $\nu_\mu$ CC (red) and $\nu$ NC (black) MonteCarlo sample of interactions in ICARUS T600. The average values refer to the entire spectrum up to 400 GeV.

2) $\nu_e$ CC events are recognized by a primary electron building up a highly ionizing electromagnetic shower and 34 GeV energy deposition on average;
3) $\nu$ NC interactions are the less energetic events with an average visible energy of 8 GeV, which can be clustered in few meters of detector.

**7.1 The PMT trigger system**

The main ICARUS T600 trigger for detecting CNGS beam related events required the coincidence of the PMT sum signal (see Section 3.2) in at least one of the four TPC chambers with a 60 µs gate (see Section 5) opened in correspondence of the proton spill extraction delayed for the ~ 2.44 ms CERN to Gran Sasso neutrino time-of-flight. As a result, a 1.8 mHz trigger rate (~70 triggers per $10^{17}$ pot, protons on target) was almost steadily obtained during the 2011 run, slightly changing with time according to the frequency of the PMT sum signal (Figure 16). The trigger rate increased up to ~ 3.5 mHz during 2012 run after the extension of the integration time over both fast and slow scintillation light components. This improvement only affected the cosmic ray collection below 0.8 GeV of energy deposition. In the few percent of spills in which the CNGS gate is not opened due to missed or late reception of the EW packets at the LNGS site (see Section 5), neutrino interactions can be recovered among non-CNGS events adopting an offline time tagging.

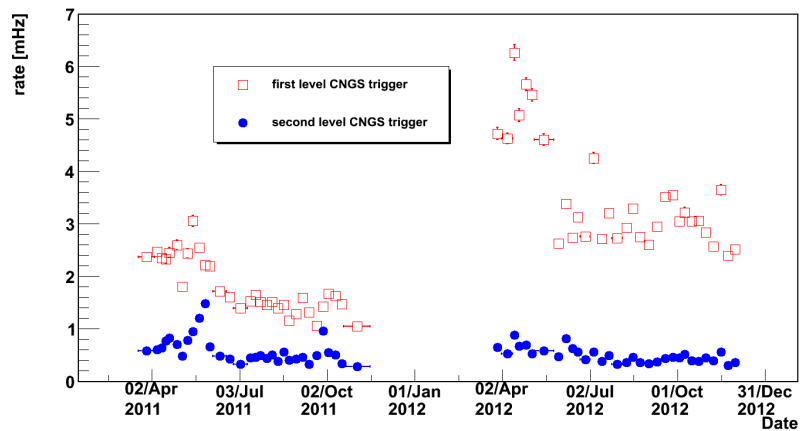

**Figure 16. Trend of the CNGS first (red) and second (blue) level trigger rate during 2011 and 2012 runs. Data have been grouped in variable-width bins, normalized to $10^{17}$ pot intensity. A relevant reduction of the electronic noise was achieved with two hardware interventions in May 2011 and May 2012.**

This trigger was assigned the highest priority by the detector Trigger Manager, i.e. always reserving at least one buffer for it when receiving any other trigger request. This allowed maintaining a negligible dead-time, ~ 0.2% during 2011 run and completely absent in 2012 thanks to further improvements introduced in the data acquisition.

Most triggered events are generated by natural radioactivity in LAr and electronic noise, resulting in a ~ 25 Hz trigger rate per chamber. An additional offline $2^{nd}$ level trigger has been added since 2011 to reject these fake signals. It consists in the software application of the DR-slw algorithm in Collection view (threshold 7 ADC counts, majority MAJ =12 and GTO > 4, see 4.1) to fully identify tracks at least ~ 40 cm long while rejecting more than 90 % of empty events (see Section 7.2). The resulting ~ 0.5 mHz trigger rate remained stable along the whole CNGS run, except for a particularly noisy period (Figure 16).

The efficiency of the PMT CNGS trigger has been estimated starting from the corresponding PMT sum signal efficiencies for each TPC, as measured with cosmic muons (see Section 3.3). The requirement of the PMT sum signal in at least one TPC per module guarantees an almost full efficiency for an energy deposition $E_{Dep} > 300$ MeV in both cryostats (Figure 17). The almost uniform PMT coverage in the East module ensures a comparable light recording for both CNGS neutrino interactions (developing along the detector longitudinal direction) and vertical muons, even at energy deposition below 1 GeV. With regard to the West module, the deployment of a single PMT row does not affect CNGS neutrino detection efficiency for energy deposition above few hundreds MeV, because the produced light largely exceeds the discrimination thresholds. This fact has been verified on a sample of $\nu_\mu$ CC and $\nu$ NC CNGS events with $E_{Dep} < 2$ GeV recorded during 2012 run, by evaluating the pulse-height of the sum signal only of PMTs in the central row of the East module. As an example, in the lowest energy NC event of the analyzed data sample ($E_{Dep} = 0.24$ GeV) 965 (697) phe have been measured in the Left (Right) Chamber. The PMT sum signal of the central row amounts to 149 (209) phe in the Left (Right) Chamber, well above the 65 phe threshold required to trigger CNGS events in the West module.

As a consequence, the PMT sum signal trigger efficiency measured with cosmic rays in 2012 can be extended to CNGS neutrino interactions below 1 GeV in both modules (Figure 17). For higher energy depositions, the PMT CNGS trigger can be considered fully efficient in the whole detector volume.

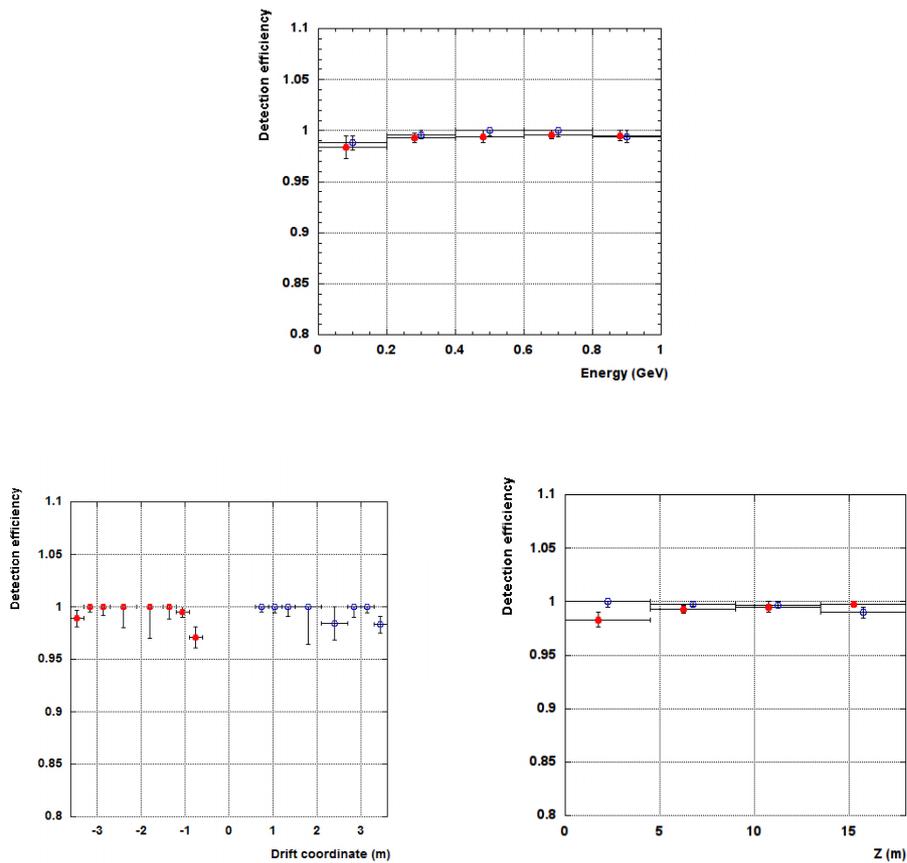

**Figure 17. Top: Efficiency of the OR of PMT sum signals in the two TPCs of West (red circles) and East (blue circles) modules, as a function of deposited energy for cosmic ray muons. The associated systematic uncertainty ranges from 3 (1) % in the lowest energy bin for the West (East) module down to a negligible level in the highest energy bin for both modules. Bottom: detection efficiency as a function of the drift**

**coordinate (left) and of the longitudinal coordinate (right) for comic ray muons. The represented error bars account only for the statistical error.**

### 7.2 The S-Daedalus trigger system

The second complementary trigger based on DR-slw algorithm applied to charge signals on the TPC wires was setup and gradually introduced starting from the 2011 run. This independent trigger allowed also qualifying the PMT trigger on the basis of a "minimum bias" request, i.e. the presence of a small track in the TPCs.

During 2011, the CNGS-gate signal was used to collect the full drift volume (1 ms drift time) in absence of PMT trigger signal. These events were then filtered to search for charge deposition, by applying the DR-slw algorithm to Collection wire signals, where the signal to noise ratio is more favorable. The algorithm parameters have been set relying on a preliminary analysis of a 2010 data sample, consisting of 109 neutrino interactions and 389 muons produced by neutrinos interacting in the rock. With 7 ADC counts threshold for single hit detection, 25 μs PEAK stretching and MAJ=12 majority, all these CNGS events were selected with $10^{-3}$ residual noise by requiring at least 7 GTO signals, corresponding to 76 cm TPC wire occupancy (Figure 18). The electronic upgrade and detector maintenance held in January 2011 allowed reducing this minimum occupancy requirement to 5 GTO (~ 40 cm), still with negligible residual noise.

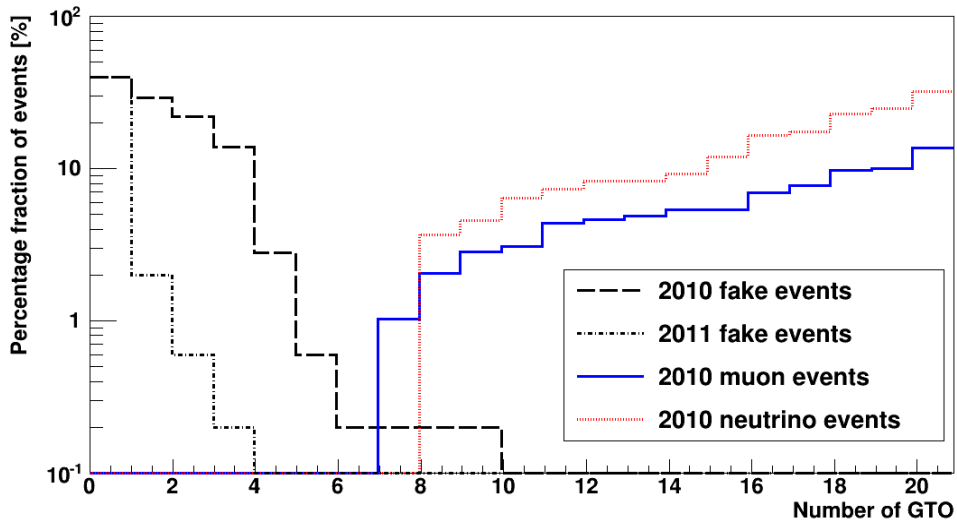

**Figure 18. Results of the DR-slw algorithm applied to the 2010 dataset. The distribution shows the number of events that would have been lost as a function of the minimum GTO number required. For comparison the noise level measured in 2010 and 2011 with empty events acquired with random triggers is also reported (black dashed lines).**

This additional trigger system was operated in steady conditions from May, 12$^{th}$ to October, 8$^{th}$ 2011, collecting a $2.5 \times 10^{19}$ pot event statistics. A $10^{19}$ pot subsample of the overall statistics has been analyzed. Globally $0.63 \times 10^{6}$ spills have been triggered, out of which 5479 passed the DR-slw software filter and 1371 contained a physical event, corresponding to a ~ 150 rejection factor of noise events. In addition to the 1074 muons and 294 neutrinos recorded with the PMT primary trigger, 6 muons from neutrinos interacting in the rock, 6 cosmic rays in spill and 1 residual of neutrino interaction outside the LAr active volume were selected only by triggering on the TPC charge deposition. This result proved the almost full PMT trigger efficiency for neutrino interactions in the detector active volume and muons from external interactions.

In the 2012 CNGS run the SD trigger system came into operation replacing the corresponding software procedure. The Trigger Control Logic was programmed to require at least 1 GTO signal in Collection view within the CNGS beam gate in absence of PMT trigger signal. The required number of GTO signals in the 2$^{nd}$ level offline filter was gradually reduced from 5 to 1 GTO (see Table 1) reaching the minimum trigger condition of a 4 cm track (1 GTO) in the whole T600 detector. The resulting SD trigger rate was steadily below 1 mHz with the deployment of an automatic veto (Figure 12) that excludes possible TPC noisy regions (Figure 19).

This CNGS-type trigger signal was assigned the second priority in the general trigger logic; the associated dead-time was steadily ~ 2 % during 2011 run, dropping down to less than 0.1 % in 2012 after the DAQ improvements.

**Table 1. Expected MonteCarlo detection efficiency of the SD trigger system for CNGS events with the parameter configurations of the 2012 run. A 125 μs signal stretching has been adopted, except for the $Q_{thr}$=7 MAJ=12 and $Q_{thr}$=8 MAJ=10 configurations, for which a 25 μs has been used. Average values refer to the related pot statistics of each parameter configuration.**

|  |  | $\nu_\mu$CC | $\nu$ NC | $\nu_e$CC |
|---|---|---|---|---|
| GTO > 0 | $Q_{thr}$=8 MAJ=10 | 99.6 ± 0.1 | 94.7 ± 0.2 | 99.5 ± 0.1 |
| GTO > 1 | $Q_{thr}$=7 MAJ=10 | 99.2 ± 0.1 | 90.7 ± 0.3 | 98.9 ± 0.1 |
|  | $Q_{thr}$=8 MAJ=10 | 99.2 ± 0.1 | 92.2 ± 0.3 | 99.0 ± 0.1 |
|  | $Q_{thr}$=7 MAJ=12 | 99.0 ± 0.1 | 90.9 ± 0.3 | 98.8 ± 0.1 |
| GTO > 4 | $Q_{thr}$=8 MAJ=10 | 97.1 ± 0.2 | 85.2 ± 0.4 | 97.0 ± 0.2 |
| AVERAGE |  | 98.8 ± 0.1 | 91.2 ±0.2 | 98.6 ±0.1 |

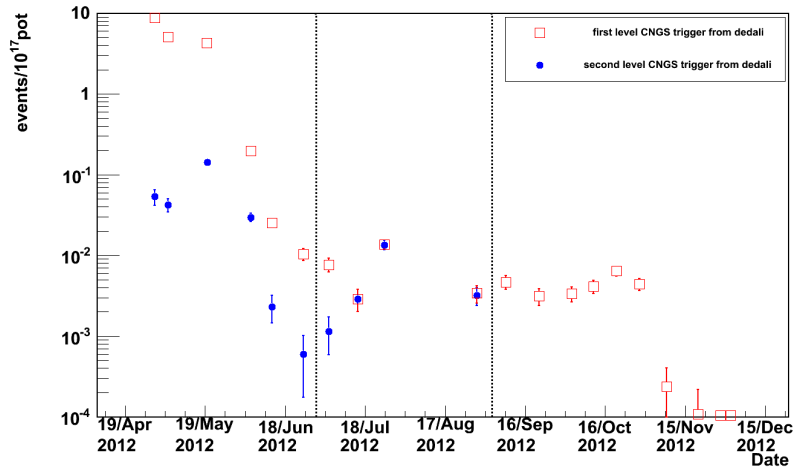

**Figure 19. SD trigger rate during 2012 CNGS run, both for 1$^{st}$ and 2$^{nd}$ level. The progressive reduction of the noise conditions in the detector allowed reducing the number of required GTO signals in the 2$^{nd}$ level trigger down to 1 GTO since the beginning of September. The 2$^{nd}$ level trigger was redundant in the last period so it was switched off in that time.**

The expected efficiency of the SD CNGS trigger has been evaluated on the $10^4$ $\nu_e$ CC, $\nu$ NC and $\nu_\mu$ CC MC sample (Table 1). The fiducial volume was selected by requiring at least 15 cm and 5 cm distance of interaction vertex from the upstream and downstream borders of the active volume of the detector, respectively, and a minimum 1.5 cm gap from each TPC border in the other directions. The actual live-time of the SD trigger system has been properly taken into account, excluding the few percent noisy boards as measured during the 2012 CNGS run.

The GTO > 0 configuration ensured full efficiency on CC interactions for both $\nu_\mu$ and $\nu_e$, while the corresponding efficiency on NC almost reached 95 %. Globally, accounting for all the SD

– 18 –

parameter configurations, efficiencies of ~ 99 % and 91 % were obtained, respectively for CC and NC events. A full trigger efficiency is guaranteed over the whole energy spectrum down to ~ 0.5 GeV for both GTO > 0 and GTO > 1 (Figure 20) while the corresponding energy deposition value increases to ~ 2 GeV (4 GeV) in $\nu_\mu$ ($\nu_e$) interactions if at least 5 GTO are required.

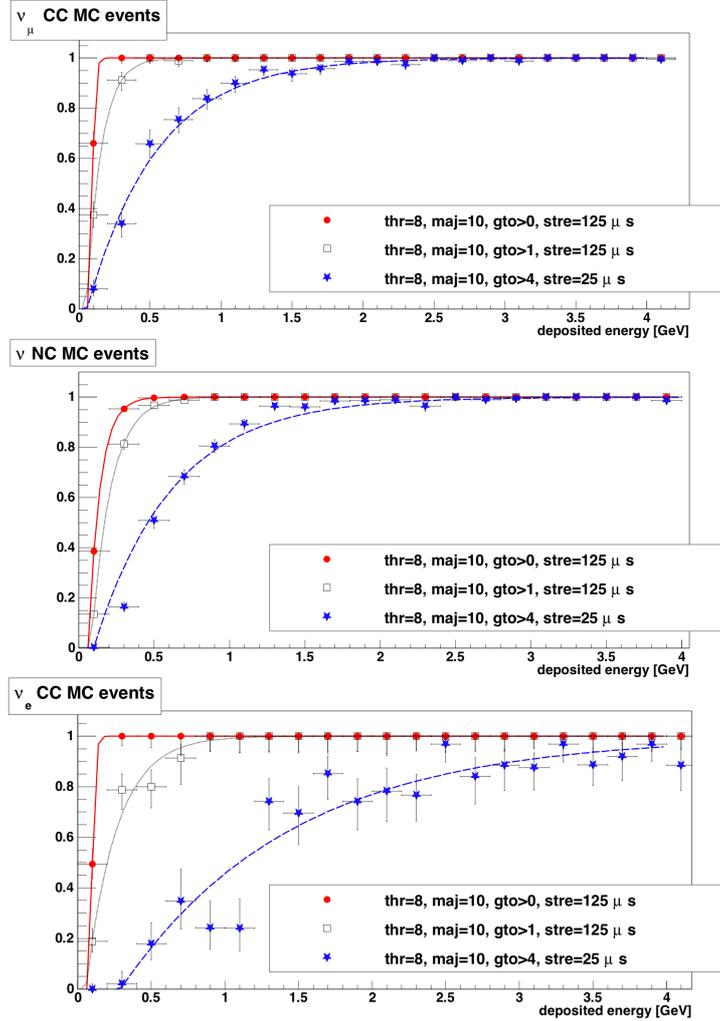

**Figure 20. SD CNGS trigger efficiency for $\nu_\mu$ CC (top), $\nu$ NC (center) and $\nu_e$ CC (bottom) MonteCarlo events as a function of the deposited energy, for three parameter configurations. The effective live-time of each SD board has been accounted for in the analysis.**

The SD trigger inefficiency at lowest energies is related to events clustered in a small fraction of the detector, where particles travel almost along the Collection wire direction with poor wire occupancy (Figure 21). This drop of performace can be recovered, in future exploitations, optimizing the majority level of the wires hit in Collection view and with the addition of SD in Induction2 view.

The evaluation of the CNGS PMT trigger performance, relying on the implementation of the SD system as a reference (Section 7.1), was not affected by this inefficiency down to 100 MeV deposited energy (Figure 22). Indeed, the amount of light collected by the PMT system from the sample of tracks at least 4 cm long selected with "minimum bias" GTO > 0 request (Section 3.3), does not depend on the track orientation in the LAr volume.

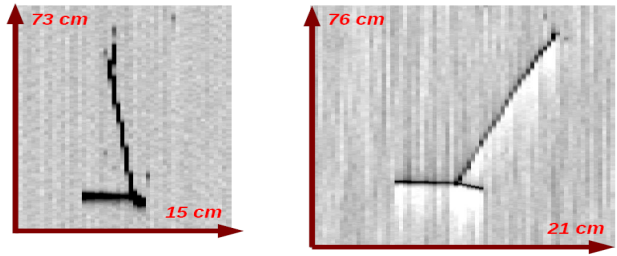

**Figure 21.** Example of a MC NC interaction of 174 MeV (deposited energy) not recognized by SD system with MAJ = 10 and GTO = 1 in Collection view (left). The 16 hit wires are shared between two adjacent 16 wire group. The Induction2 view is also shown (right).

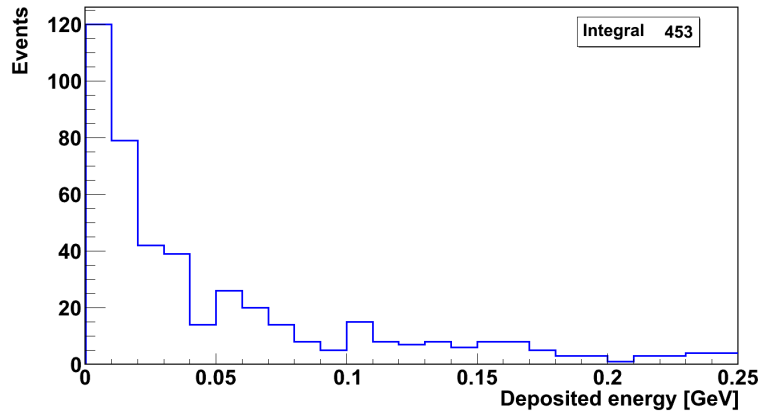

**Figure 22.** Distribution of the deposited energy of ν NC MC interactions, not recognized by SD system with MAJ = 10 and GTO = 1, from a total of $10^4$ generated events. Events below ~ 0.1 GeV are due to isolated converted gammas, neutrons and protons. The remaining is due to tracks projected over few Collection wires or shared between consecutive groups of 16 wires.

## 8. Final discussion of ICARUS trigger for CNGS

Two different trigger systems based on the detection of scintillation light and ionization charge produced by charged particles in LAr have been realized for the ICARUS T600 detector. They exploit the PMT arrays and the new S-Daedalus FPGA boards, spanning few orders of magnitude in event energy deposition.

The combined analysis of the performance of the PMT and S-Daedalus independent trigger systems demonstrated an almost full PMT trigger efficiency for CNGS neutrino events above 300 MeV energy deposition on the full T600 active volume, remaining ~ 98.5% efficient down to 100 MeV during 2012. The stability of the trigger system was verified within the measurement uncertainty, comparing different data sets collected during the CNGS 2012 run.

The 2011 CNGS run was performed with a similar trigger configuration based on the PMT sum signals within the CNGS beam gate but before the PMT electronics upgrading. No recording of the PMT signals was available, preventing a trigger efficiency evaluation. Nevertheless, accounting for the stability of the PMT system during data taking, some information can still be deduced by comparing the cosmic muon spectra collected in 2011 and in 2012 by triggering on the coincidence of the PMT sum signals from the two TPCs in the same module (a more stringent request with respect to the single PMT sum signal used in the CNGS trigger). Due to the huge photon emission by charged particle in LAr, no major differences in trigger efficiency have been measured in both modules at least down to 1 GeV deposited energy (Figure 23) [18]. Since the trigger efficiency demonstrated to be

– 20 –

uniform with respect to polar angle, this result can be extended to CNGS triggers as well. Furthermore, the direct comparison of the 2011 PMT trigger performance for CNGS events with the DR-slw algorithm, applied to a set of data recorded acquiring a full drift volume for every CNGS spill, proved an almost full PMT trigger efficiency both for neutrino interactions and muons from external interactions (Section 7.2). The direct data processing for two periods of 2011 and 2012 in which the CNGS facility run in nominal conditions, showed an average rate of $3.4 \pm 0.1$ neutrino interaction events per $10^{17}$ pot inside LAr-TPC in 2011 well in agreement, within the errors, with the corresponding $3.5 \pm 0.1$ events per $10^{17}$ pot observed in 2012.

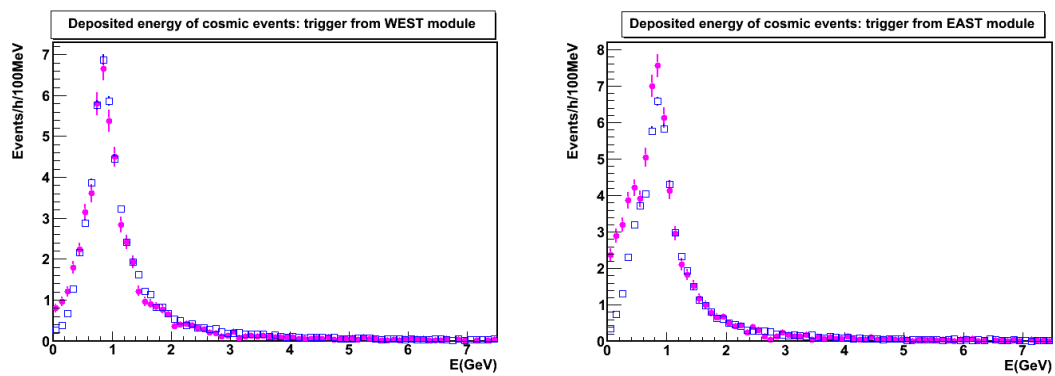

**Figure 23.** Comparison between the differential cosmic ray spectra in the West (left) and East (right) modules collected in July 2011 (blue open squares) and in March 2012 (pink full circles), after the PMTs improvements carried out from December 2011 to March 2012.

The obtained results demonstrate that the implemented ICARUS T600 trigger system has been effective for a wide range of event energies throughout the CNGS run in a stable way (Table 2), setting a benchmark for forthcoming LAr-TPC detectors.

**Table 2.** CNGS trigger efficiencies of the two ICARUS T600 modules. In 2011 data taking the trigger efficiency is expected to smoothly degrade below 1 GeV energy deposition as demonstrated by the c-rays recorded spectra (Figure 23). The error on the measurement for the 2012 event sample with deposited energy in the $0.1 \div 0.3$ GeV energy range is of the order of 2%, with similar statistics and systematics contributions. In all other cases the precision can be estimated better than 1% accounting also for the inhomogeneous PMT coverage with respect to the spatial distribution of the analyzed track sample.

|  | 2011 | 2012 | |
|---|---|---|---|
|  | $E_{Dep} > 1$ GeV | $E_{Dep}\ 0.1 \div 0.3$ GeV | $E_{Dep} > 0.3$ GeV |
| **West cryostat** | 99.5 % | 98.4 % | 99.6 % |
| **East cryostat** | 100 % | 98.8 % | 99.8 % |

## Acknowledgements

The successful results reported in this paper could not have been achieved without the effort of the electronics services of INFN-LNGS, INFN Padova and INFN Pavia that contributed to the implementation and operation of the ICARUS T600 trigger system. In particular the contributions of F. Calaon, S. Marchini, M.C. Prata and P.G. Zatti are acknowledged.

The Polish groups acknowledge the support of the National Science Center: Harmonia (2012/04/M/ST2/00775) and Preludium (2011/03/N/ST2/01971) funding schemes.## References